\documentclass[11pt]{article}

\usepackage{amssymb,latexsym, amsmath}
\usepackage{graphicx}

\makeatletter
\@addtoreset{figure}{section}
\def\thefigure{\thesection.\@arabic\c@figure} \def\fps@figure{h, t}
\@addtoreset{table}{bsection}
\def\thetable{\thesection.\@arabic\c@table} \def\fps@table{h, t}
\@addtoreset{equation}{section}

\makeatother

\advance \topmargin by -\headheight
\advance \topmargin by -\headsep
\evensidemargin \oddsidemargin
\newtheorem{thm}{Theorem}[section]
\newtheorem{prop}[thm]{Proposition}
\newtheorem{lem}[thm]{Lemma}

\newtheorem{cor}[thm]{Corollary}

\newfont{\tenbi}{cmbxti10}

\def\one{\mathbf 1}

\def\D{\mathfrak D}

\def\P{\mathcal P}
\def\S{S}

\def\H{\mathcal H}
\def\S{\mathfrak S}
\def\L{\mathfrak L}

\def\n{\mathfrak n}
\def\b{\mathfrak b}
\def\O{\mathfrak O}

\def\B{\mathbf B}

\begin{document}

\title{Inverse moment problem for elementary co-adjoint orbits}
\author {Leonid Faybusovich
\thanks{Research partially supported by the NSF grant  DMS98-03191}
\\
Department of Mathematics\\
University of Notre Dame\\
Notre Dame, IN 46556 \\
{\footnotesize Leonid.Faybusovich.1@nd.edu}\\
\and
Michael Gekhtman\\
Department of Mathematics\\
University of Notre Dame\\
Notre Dame, IN 46556 \\
{\footnotesize Michael.Gekhtman.1@nd.edu}
}
\maketitle

\begin{abstract}
We give a solution to the inverse moment problem
for a certain class of Hessenberg and symmetric
matrices related to integrable lattices of Toda type.
\end{abstract}

\section{Introduction}

This paper continues our previous work \cite{fg1,fg2},
where we dealt with the family of integrable Hamiltonian systems
in $\mathbb{R}^{2n}$
parametrized by index sets $I=\{i_1,\ldots,i_k\ : 1< i_1< \ldots< i_k=n\}$
and generated by Hamiltonians
\begin{equation}
{\displaystyle
H_I(Q,P) =\frac{1}{2} \sum_{i=1}^n P^2_i\quad + \sum_{1\le i < n; i\ne
i_1,\ldots,i_{k-1}}^n P_i e^{Q_{i+1} - Q_i}\ +\
\sum_{j=1}^{k-1}e^{Q_{i_j+1} - Q_{i_j}}\ .
}
\label{hamIexp}
\end{equation}
This family contains (after an appropriate coordinate changes)
such  important integrable systems as the standard and relativistic
Toda lattices, Volterra lattice and  lattices of the ~Ablowitz-~Ladik hierarchy.

In \cite{fg1}, we argued that the full Kostant-Toda flows on Hessenberg
matrices provide a convenient framework to study systems generated by
(\ref{hamIexp}). Namely, each of these systems possesses a Lax representation
with a Lax operator given by an $n\times n$  upper Hessenberg matrix $X_I=X_I(Q,P)$
that belongs
to a certain $(2n-2)$-dimensional co-adjoint orbit of the upper triangular group.
This orbit is determined by $I$ and the value of $Tr(X_I)= \sum_{i=1}^n P_i$.
Furthermore, $Tr(X_I^2)=H_I$ and Hamilton equations of motion are equivalent
to the Toda flow ${\dot X}_I = [X_I, (X_I)_{\le 0}]$, where $(A)_{\le 0}$
denotes the lower triangular part of a matrix $A$.
The orbit  contains a dense open set of elements that admit a factorization
into a product of elementary bi-diagonal matrices. Written in terms of parameters
of this factorization, equations of motion become a particular case of the
constrained KP lattice studied in \cite{sur3}.

Each of the systems described above can be linearized via the Moser map
$X \to m(\lambda, X)= ((\lambda\one-X)^{-1} e_1,e_1)=
\sum_{j=0}^\infty\frac{s_j(X)}{\lambda^{j+1}}$ where $e_1=(1,0,\ldots,0)$ and
$s_j(X)=(X^{j} e_1,e_1)$ (see \cite{moser,b,dlnt,bf}).
Moreover, as we have shown in \cite{fg2} the Moser map is very useful in establishing
a multi-Hamiltonian structure of these systems. However, explicit formulas
for the inverse of the Moser map seem to be known only in two cases.
If $X_I$ is tri-diagonal ($I=\{ 2,\ldots,n\}$) they give a solution in terms
of Hankel determinants to the classical
finite-dimensional moment problem (see, e.g.  \cite{akh}). In the "opposite"
case that corresponds to the relativistic Toda lattice ( $I=\{ n\}$) the solution
to the inverse problem is given in terms of Toeplitz determinants constructed from
the moments $s_j(X_I), j=-n,\ldots, n$ (see \cite{kmz}).

The main purpose of this paper is to give a solution to the inverse problem for
any $I$. Explicit formulas (given, once again, in terms of Toeplitz determinants)
are obtained in sect. 3.

As in the Hessenberg case, the inverse problem
for elementary co-adjoint orbits
in the symmetric case was
previously studied for $I=\{ 2,\ldots,n\}$ and $I=\{ n\}$. The latter case
corresponds to peakons solutions of the shallow water equation and
was recently comprehensively studied in \cite{bss}. We treat the symmetric
case in sect. 4 for an arbitrary $I$.

We would like to express our gratitude to Yu. Suris for valuable comments
and suggestions.

\section{Elementary Orbits in the Hessenberg Case}

Everywhere below we denote by $e_{jk}$ an elementary matrix
$(\delta_\alpha^i \delta_\beta^k)_{\alpha, \beta=1}^n$ and by
$e_j$ a column vector $(\delta_\alpha^j)_{\alpha=1}^n$ of the
standard basis in $\mathbb R^n$.

Denote by $J$ an $n\times n$ matrix with $1$s on the first sub-diagonal and $0$s everywhere else. Let $\b_+, \n_+, \b_-, \n_-$ be, resp.,
algebras of upper triangular, strictly upper triangular, lower triangular and strictly lower
triangular matrices. Denote by $\H$ the set $J + \b_+$ of upper Hessenberg matrices.

For any matrix $A$ we write its decomposition into a sum of lower triangular and strictly upper triangular matrices as
\begin{equation}\nonumber
A=A_- + A_0 + A_+
\end{equation}
and define $A_{\ge 0}= A_0 + A_+,A_{\le 0}= A_0 + A_-,\ A_{sym}= A_+ + A_0 + A^T_+$.

A linear Poisson structure on $\H$ is
obtained as a pull-back of the ~Kirillov-Kostant
structure on $\b_-^*$, the dual of $\b_-$, if one identifies $\b_-^*$
and $\H$ via the trace form. A Poisson bracket of two functions $f_1, f_2$ on
$\H$ then reads
\begin{equation}
\{f_1, f_2\} (X)=
\langle X, [(\nabla f_1 (X))_{\le 0}, (\nabla f_2 (X))_{\le 0}] \rangle\ ,
\label{brack1}
\end{equation}
where we denote by $\langle X, Y\rangle$ the trace form $\mbox{Trace}(XY)$
and gradients are computed w.r.t. this form.
Symplectic leaves of the bracket (\ref{brack1}) are
orbits of the coadjoint action of the group $\B_-$ of lower triangular invertible matrices:
\begin{equation}
\O_{X_0}=\left \{ J + (\mbox{Ad}_n X_0)_{\ge 0}\ : n\in \B_-\right \}\ .
\label{coad}
\end{equation}

Following \cite{fg1}, consider a family of orbits whose members
are parameterized by increasing sequences of natural
numbers
$I=\{i_1,\ldots,i_k\ : 1< i_1< \ldots< i_k=n\}$. To each sequence $I$ there
corresponds a $1$-parameter family of $2 (n-1)$-dimensional coadjoint
orbits
\begin{equation}
{\displaystyle
M_I=\cup_{\nu\in \mathbb R} \O_{X_I+ \nu \one}\subset \H\ ,
}
\label{M_I}
\end{equation}
where
\begin{equation}
{\displaystyle
X_I=e_{1 i_1} + \sum_{j=1}^{k-1}e_{i_j i_{j+1}} + J\ .
}
\label{X_I}
\end{equation}
We call orbits $\O_{X_I+ \nu \one}$ {\it elementary}.

The set $M'_I$ of elements of the form
\begin{equation}
{\displaystyle
X=(J+D) (\one - C_k)^{-1} (\one - C_{k-1})^{-1}\cdots (\one -
C_1)^{-1},
}
\label{factorI}
\end{equation}
where $D=\mbox{diag} (d_1,\ldots, d_n)$
\begin{equation}
{\displaystyle
C_j= \sum_{\alpha=i_{j-1}}^{i_j-1} c_\alpha e_{\alpha,\alpha+1}\ ,
}
\label{Cj}
\end{equation}
is dense in $M_I$.

The following formulae express      entries $x_{lm}\ (l<m)$ of $X$ in terms of $c_i, d_i$ (see \cite{fg1}) :
\begin{equation}
x_{lm}=d_l u_{lm} + u_{l-1,m}=\left \{ \begin{array}{cc}
(d_l +c_{l-1}) c_l\cdots c_{m-1}, &\ i_{j-1} < l < m \le i_j\\
d_{i_{j-1}}c_{i_{j-1}}\cdots c_{m-1}, &\ i_{j-1} = l < m \le i_j\\
d_l + c_{l-1}, &\ l=m\\
0, & \ \mbox{otherwise}
\end{array} \right .
\label{xlm}
\end{equation}
(Here $c_0=0$.)





Define a sequence $\epsilon_1,\ldots \epsilon_n$ by setting
\begin{equation}
\epsilon_i=\left \{ \begin{array}{cc} 0 & \mbox{if}\  i=i_j\
\mbox{for some}\  0< j \le k\\ 1 & \mbox{otherwise}\end{array}
\right .\ . \label{eps}
\end{equation}
Then $X$ can also be written
as\footnote{This was suggested to us by Yu. Suris \cite{surpri}}
\begin{equation}
{\displaystyle
X=(J+D) (\one + U_1) (\one - U_2)^{-1},
}
\label{factorIU}
\end{equation}
where
\begin{equation}
{\displaystyle U_1= \sum_{\alpha=1}^{n-1} (1- \epsilon_\alpha)
c_\alpha e_{\alpha,\alpha+1}\ , U_2=\sum_{\alpha=1}^{n-1}
\epsilon_\alpha c_\alpha e_{\alpha,\alpha+1} } \label{U12}
\end{equation}

In what follows, we will also use  a sequence of integers
$(\nu_i)_{i=1}^{n}$ defined by
\begin{equation}
\nu_i =i (1-\epsilon_i) -\sum_{\beta=1}^{i -1} \epsilon_\beta
\label{nu}
\end{equation}
It is easy to check that
\begin{equation}
\nu_i=\left \{ \begin{array}{cc} j & \mbox{if}\  i=i_j\ \mbox{for
some}\  0< j \le k\\ -\sum_{\beta=1}^{i -1} \epsilon_\beta &
\mbox{otherwise}\end{array} \right .\ . \label{nunu}
\end{equation}

It follows from (\ref{nunu}) that
\begin{itemize}
\item[(i)] sequences $(\epsilon_i)$ and $(\nu_i)$ uniquely
determine each other.
\item[(ii)] If we define, for every $i\in \{1,\ldots,n\}$, a set
$N_i=\{\nu_\alpha\ : \ \alpha=1,\ldots,i \}$ and a number
$k_i=\max \{ j: i_j \le i\}=i-\sum_{\beta=1}^i \epsilon_\beta$,
then
\begin{equation}
N_i= \{ k_i -i+1,\ldots, k_i-1,k_i \}= \{1 - \sum_{\beta=1}^i
\epsilon_\beta, \ldots,  i-\sum_{\beta=1}^i \epsilon_\beta \}
\label{interval}
\end{equation}
\end{itemize}

\section{Solution of the Inverse Problem}

Assume that all $d_i\ne 0$ and then define the moment sequence
$S=(s_i, i\in \mathbb Z)$ of $X$:
\begin{equation}
s_i=s_i(X)=e^T_1 X^i e_1 . \label{moments}
\end{equation}

Our goal is to express coefficients $c_i, d_i$ in terms of $S$.
In fact, only a segment $s_{k+1-n}, \ldots, s_{n+k}$ will be needed.

Let $p(\lambda)=(p_0 (\lambda):=1, p_1 (\lambda), \ldots,
p_{n-1} (\lambda))$ be a solution of the truncated eigenvalue problem
\begin{equation}{\displaystyle
(p(\lambda) X)_i=\sum_{\alpha=1}^i x_{\alpha i}
p_{\alpha-1}(\lambda) + p_i(\lambda)=\lambda p_{i-1}(\lambda), \
i=1,\ldots, n-1 }
\label{eigen}
\end{equation}
or, equivalently,
\begin{equation}
\left (p(\lambda) (J+D) (\one + U_1)\right )_i=\lambda \left ((\one - U_2) p\right )_i(\lambda), \ i=1,\ldots, n-1 \ .
\label{eigen2}
\end{equation}
Clearly, such solution exists for every $\lambda$ and is uniquely defined.
Moreover, each $p_i(\lambda)$ is a monic polynomial of degree $i$.
We can re-write (\ref{eigen2}) as a $3$-term recursion for polynomials
$p_i(\lambda)$ :
\begin{equation}
p_{i+1}(\lambda) + b_{i+1} p_i(\lambda) + (1-\epsilon_{i}) a_i p_{i-1}(\lambda)
=\lambda \left ( p_i(\lambda) - \epsilon_{i} c_{i} p_{i-1}(\lambda)  , \right ) \ i=0,\ldots, n-1 \ ,
\label{3-term}
\end{equation}
where
\begin{equation}
b_i= d_i + (1-\epsilon_{i-1}) c_{i-1},\  a_i = d_{i}c_{i}\ .
\label{ab}
\end{equation}

\begin{lem}
For any upper Hessenberg matrix $X$, polynomials $p_i(\lambda)$
defined by (\ref{eigen}) satisfy
\begin{equation}
p_i(X) e_1= e_{i+1},\  i=0,\ldots, n-1 .
\label{basis}
\end{equation}
\label{2.1}
\end{lem}

\noindent{\bf Proof.} Since
$$ {\displaystyle X e_i= \sum_{\alpha=1}^i x_{\alpha i}
e_{\alpha} + e_{i+1} }$$ and, by (\ref{eigen}),
$$ X p_{i-1} (X) = \sum_{\alpha=1}^i x_{\alpha i}
p_{\alpha-1}(X) e_1 + p_i(X) e_1\ ,$$ one concludes that sequences
of vectors  $(e_i)_{i=1}^n$ and $(p_{i-1} (X) e_1)_{i=1}^n$ are
defined by the same recurrence relations. \hfill $\Box$

\begin{lem}
For $i=1,\ldots, n-1$, a  subspace  $\L_i$ generated by vectors
$(e^T_\alpha)_{\alpha=1}^i$ coincides with a subspace generated
by vectors $(e^T_1 X^{\nu_\alpha})_{\alpha=1}^i $, where
$\nu_\alpha $ are defined in (\ref{nu}). In other words, for some
constants $\gamma_i\ne0$
\begin{equation}
{\displaystyle
\gamma_{i} e^T_{i} =  \left \{ \begin{array}{cc} e^T_1 X^{j} & \mbox{if}\ i= i_j\\
e^T_1 X^{-\sum_{\beta=1}^{i-1} \epsilon_\beta} & \mbox{otherwise}
\end{array} \right . \ (\mbox{mod}\ \L_{i-1}) ) . } \label{flag1}
\end{equation}
Moreover, for $i=2,\ldots, n$,
\begin{equation}
{\displaystyle \gamma_{i} = e^T_1 X^{i (1-\epsilon_i)
-\sum_{\beta=1}^{i -1} \epsilon_\beta} e_{i} =
= (-1)^{\epsilon_i (i-1)} c_1 d_1 \frac{\prod_{\beta=2}^{i-1}
d_\beta^{1-\epsilon_\beta}c_\beta}{(d_1\cdots
d_{i})^{\epsilon_i}} \ . } \label{gamma}
\end{equation}
\label{lemmaflag}
\end{lem}

\noindent{\bf Proof.}
For any $X$ given by (\ref{factorI}), (\ref{factorIU})  ,consider
an upper triangular matrix
$$
V=D (\one - C_k)^{-1} (\one - C_{k-1})^{-1}\cdots (\one -
C_1)^{-1}=D (\one + U_1) (\one - U_2)^{-1}\ .
$$
(If $D$ is invertible, then $V$ is an upper triangular factor in
the Gauss factorization of $X$.)

By (\ref{Cj}), $e_l^T C_j=0$ for $l\le i_{j-1}$ and $l\ge i_j$.
Thus, for $j=0,\ldots,k$,
\begin{eqnarray}\nonumber
&e_{i_{j-1}}^T V=d_{i_{j-1}}e_{i_{j-1}}^T (\one - C_k)^{-1} \cdots
(\one - C_1)^{-1}=d_{i_{j-1}}e_{i_{j-1}}^T (\one - C_j)^{-1} (\mbox{mod}\ \L_{i_j-1})\\
\nonumber &= d_{i_{j-1}}c_{i_{j-1}}\cdots c_{i_{j}-1}e^T_{i_{j}}\
(\mbox{mod}\ \L_{i_j-1}) \ .
\end{eqnarray}
Similar argument shows that $ e_l^T V \in  \L_{i_j-1}$ for $l <
i_{j-1}$. This implies
\begin{eqnarray}\nonumber &e_{1}^T V^j=
e_{1}^T V^j\ (\mbox{mod}\ \L_{i_j-1})= \left
(\prod_{\beta=0}^{j-1} d_{i_{\beta}}c_{i_{\beta}}\cdots
c_{i_{\beta+1}-1}\right ) e_{i_{j}}\ (\mbox{mod}\ \L_{i_j-1})\\ &
= \left ( c_1 d_1 \prod_{\beta=2}^{i_j-1} c_\beta
d^{1-\epsilon_\beta}_\beta \right ) e^T_{i_{j}}\ \ (\mbox{mod}\
\L_{i_j-1}). \label{aux2}
\end{eqnarray}
On the other hand,  for $l\in \{i_{j-1},\ldots,i_j-1\}$, define
$m\ge 0$ to be a number such that $ l+\beta=i_{j+\beta -1}$ for
$\beta=1,\ldots, m$ and $l+m+1 < i_{j+m}$. In other words,
$l+m+1$ is the smallest index greater than $l$ that does not
belong to the index set $I$. Then
\begin{equation}
e_l^T V^{-1}= e_l^T(\one - C_j)\cdots(\one -
C_k)D^{-1}=
\left ((-1)^{m+1} c_l\cdots c_{l+m} d_{l+m+1}^{-1}\right )
e_{l+m+1}^T (\mbox{mod}\ \L_{l+m})\ . \label{aux3}
\end{equation}

Let us denote by $J=\{ l_1 < \ldots < l_{n-k-1}\}$ the set
$\{1,\ldots,n\} \backslash I$. Then (\ref{aux3}) implies
\begin{eqnarray}\nonumber
&e_1^T X^{-\alpha}=
e_1^T V^{-\alpha}\ (\mbox{mod}\ \L_{l_\alpha-1})\\ &= \left (
(-1)^{l_\alpha-1} c_1\cdots c_{l_\alpha-1} d^{-1}_{l_1}\cdots
d^{-1}_{l_\alpha}\right ) e_{l_\alpha}^T\ (\mbox{mod}\
\L_{l_\alpha-1})\  . \label{aux4}
\end{eqnarray}
Note now that $l_\alpha=i$ if and only if
$\alpha=\sum_{\beta=1}^{i-1} \epsilon_i$. Furthermore,
$d^{-1}_{l_1}\cdots d^{-1}_{l_\alpha}= \prod_{\beta=2}^i
d_\beta^{-\epsilon_\beta}$. Thus, one can re-write (\ref{aux4}) as
\begin{equation}
e_1^T V^{-\sum_{\beta=1}^{i-1} \epsilon_i}=(-1)^{i-1}
\prod_{\beta=1}^{i-1} c_\beta d_{\beta+1}^{-\epsilon_{\beta+1}}
e_i^T\ (\mbox{mod}\ \L_{i-1})= (-1)^{i-1} c_1 d_1 {\frac{
\prod_{\beta=2}^{i-1} c_\beta
d_{\beta}^{1-\epsilon_{\beta}}}{d_1\cdots d_i}} \ (\mbox{mod}\
\L_{i-1}) . \label{aux5}
\end{equation}
Combining (\ref{aux2}), (\ref{aux5}) with (\ref{nunu}), one
concludes that \begin{equation} e_1^T X^{\nu_i}= \gamma_i e_{i}^T
(\mbox{mod}\ \L_{i-1})\  , \label{aux6}
\end{equation} where
$\gamma_i$ is defined by (\ref{gamma}).
This implies the
statement of the lemma. \hfill $\Box$

\begin{cor}
\begin{equation}
e^T_1 X^\alpha p_i(X)e_1 =0,\  1 - \sum_{\beta=1}^{i}
\epsilon_\beta\le \alpha\le i - \sum_{\beta=1}^{i} \epsilon_\beta
\label{ortho}
\end{equation}
\label{cor}
\end{cor}

\noindent{\bf Proof.} By (\ref{flag1}), $e_1^T X^{\nu_l}
e_{i+1}=0$ for $l=1,\ldots, i$. But, by Lemma 2.1, $e_{i+1}=p_i(X)
e_1$. Then (\ref{ortho}) follows from (\ref{interval}). \hfill
$\Box$

Define Toeplitz matrices
\begin{equation}
T^{(l)}_i=(s_{l+\alpha-\beta})_{\alpha,\beta=1}^i
\label{toep}
\end{equation}
and Toeplitz determinants
\begin{equation}
\Delta^{(m)}_i=\det T^{(i +m -
\sum_{\beta=1}^{i}\epsilon_\beta)}_i\  (m\in \mathbb Z)\ .
\label{det}
\end{equation}

Let us also define polynomials
\begin{equation}
\P^{(l)}_i(\lambda)=\det \left [\begin{array}{cccc} s_{l} &
s_{l+1} & \cdots &s_{l+i}
\\ \cdots &\cdots &\cdots &\cdots
\\
s_{l-i+1} &s_{l-i+2} &\cdots & s_{l+1}
\\
\\ 1 & \lambda
&\cdots & \lambda^{i}
\end{array}\right ]\ .
\label{polytoep}
\end{equation}
$\P^{(l)}_i$ is a determinant of the $(i+1)\times (i+1)$ matrix
obtained from $T^{(l)}_{i+1}$ by replacing the last row with $(1
, \lambda ,\ldots,  \lambda^{i})$. The following simple lemma is
reminiscent of the construction of classical orthogonal
polynomials on the real line and the unit circle (see, e.g.
\cite{akh}) and will be useful for us in what follows.

\begin{lem}
Let $X$ be invertible matrix with a moment sequence
(\ref{moments}). For $m\in \mathbb{Z}$, define a Laurent
polynomial
\begin{equation}
R(\lambda)=  \lambda^m \P^{(l)}_i(\lambda)\ . \label{laurent}
\end{equation}
Then
\begin{eqnarray}
& e_1^T R(X) X^\alpha e_1=0\quad (\alpha =l+1-m-i,\ldots,l-m)
\label{det_lm}\\
&e_1^T R(X) X^{l-m+1} e_1=(-1)^i\det T^{(l+1)}_{i+1}
\label{last}\\
& e_1^T R(X) X^{l-m-i} e_1=\det T^{(l)}_{i+1}\ . \label{first}
\end{eqnarray}

\label{polydet}
\end{lem}

\noindent{Proof.} $R(\lambda)$ can be written as
$\sum_{\beta=0}^i R_{\beta} \lambda^{m+\beta}$, where $R_{\beta}$
is equal to $(-1)^{i+\beta}$ times the minor obtained by deleting
the last row and $(\beta+1)$st column in (\ref{polytoep}). Then
\begin{equation}
e^T_1  R(X)X^\alpha e_1 =\sum_{\beta=0}^i R_{\beta} s_{m+ \alpha +
\beta}\  .
\label{R(X)}
\end{equation}
If $\alpha$ is in the range specified in (\ref{det_lm}), the
right-hand side of (\ref{R(X)}) becomes a determinant in which
two of the rows coincide. Equalities (\ref{last}), (\ref{first})
are obtained in a similar way. \hfill$\Box$

Lemmas above imply the following

\begin{prop} Assume that $\Delta_i^{(0)} \ne 0$ for $i=1,\ldots,
n$. Then polynomials defined by formulas
\begin{equation}
p_i(\lambda)=\frac{1}{\Delta^{(0)}_{i}} \P^{(i-
\sum_{\beta=1}^{i}\epsilon_\beta)}_i(\lambda)\ .
\label{poly}
\end{equation}
satisfy (\ref{eigen}, \ref{eigen2},
\ref{3-term}).
\end{prop}

\noindent{\bf Proof.} If $p_i(\lambda) =\sum_{l=0}^i p_{il}
\lambda^l$ is a solution of \ref{eigen}, \ref{eigen2},
\ref{3-term}), then by Corollary \ref{cor} and (\ref{moments}),
we have
\begin{equation}
e^T_1 X^\alpha p_i(X)e_1 =0
\quad (1 - \sum_{\beta=1}^{i} \epsilon_\beta\le \alpha\le i -
\sum_{\beta=1}^{i} \epsilon_\beta \ )\ . \label{orthomom}
\end{equation}
If $\Delta_i^{(0)} \ne 0$, Lemma \ref{polydet} implies that
(\ref{poly}) is a unique monic polynomial of degree $i$ that
satisfies (\ref{orthomom}). Therefore it  has to coincide with
the unique solution of (\ref{eigen}). This completes the proof.
\hfill $\Box$

Now we will be able to express parameters $c_i, d_i$ in terms of
Toeplitz determinants (\ref{det}). First, we  need the following
elementary lemma.

\begin{lem} Let $X$ be an arbitrary upper Hessenberg matrix
and let $D_0=1$ and $D_i \ (i=1,\ldots, n)$ denote the left upper
$i\times i$ principal minor of $X$. Then
\begin{equation}
D_i=(-1)^i p_i(0)\ ,
\label{principal}
\end{equation}
where $p_i(\lambda)$ are polynomials defined by (\ref{eigen}).
\end{lem}

\noindent{\bf Proof.} The Laplace expansion w.r.t $i$th column
leads to the following recursion for $D_i$:
$$ {\displaystyle
D_i= \sum_{\alpha=1}^i (-1)^\alpha x_{\alpha i} D_{\alpha-1} }
$$
or
$$ {\displaystyle
D_i + \sum_{\alpha=1}^i (-1)^{\alpha-1} x_{\alpha i}
D_{\alpha-1}=0\ , }
$$
which coincides with the recursion for $p_i(0)$ obtained when one
sets $\lambda=0$ in (\ref{eigen}).\hfill $\Box$

\begin{thm}
\begin{equation}
\displaystyle{
d_i =\frac{\Delta^{(1)}_{i} \Delta^{(0)}_{i-1}}{\Delta^{(0)}_{i} \Delta^{(1)}_{i-1}}\ ,\
c_i = - \frac{\Delta^{(\epsilon_{i+1})}_{i+1} \Delta^{(1-\epsilon_{i})}_{i-1}}{\Delta^{(0)}_{i} \Delta^{(1)}_i}\ .
}
\label{cd}
\end{equation}
\label{invcd}
\end{thm}

\noindent {\bf Proof.} First note that, by (\ref{poly}),
the right-hand side of (\ref{principal}) is equal to
$\frac{\Delta^{(1)}_i}{\Delta^{(0)}_{i}}$, whereas, due to
(\ref{factorI}), the left-hand side is $d_1\cdots d_i$. This
immediately implies the first of the formulae (\ref{cd}).

Secondly, it is clear from (\ref{gamma}), that parameters $c_i$
are uniquely determined by $d_i$ and $\gamma_i$. On the other
hand, (\ref{gamma}), together with (\ref{basis}), gives
\begin{equation}
\gamma_{i+1} = e^T_1 X^{\nu_{i+1}}p_i(X) e_1
\ .
\label{gammapoly}
\end{equation}
Recall that, if $\epsilon_{i+1}=1$ then $\nu_{i+1}=
-\sum_{\beta=1}^{i}\epsilon_\beta$ and $i-
\sum_{\beta=1}^{i}\epsilon_\beta=i+1-
\sum_{\beta=1}^{i+1}\epsilon_\beta$. Otherwise, $\nu_{i+1}=
i+1-\sum_{\beta=1}^{i}\epsilon_\beta=i+1-\sum_{\beta=1}^{i+1}\epsilon_\beta
$. Comparing (\ref{gammapoly}) with equalities (\ref{last}),
(\ref{first}) and then using (\ref {det}), one obtains
\begin{equation}
\gamma_{i+1}= (-1)^{i \epsilon_{i+1} } c_1 d_1 \frac{\prod_{\beta=2}^{i}
d_\beta^{1-\epsilon_\beta}c_\beta}{(d_1\cdots
d_{i+1})^{\epsilon_{i+1}}}=
(-1)^{i(1-\epsilon_{i+1})}\frac{\Delta^{(0)}_{i+1}}{\Delta^{(0)}_{i}}
\label{gammadet}
\end{equation}

Now, to finish the proof, it is enough to check that when one
substitutes expressions (\ref{cd}) into the right-hand side of
(\ref{gamma}), the result agrees with (\ref{gammadet}). It is not
hard to see that if $d_\beta, c_\beta$ are defined by (\ref{cd})
then
\begin{equation}
{d_\beta}^{1-\epsilon_\beta} c_\beta = -
\frac{\Delta^{(\epsilon_{\beta+1})}_{\beta+1} \Delta^{(0)}_{\beta-1}}
{\Delta^{(\epsilon_{\beta})}_{\beta} \Delta^{(0)}_{\beta}}\ ,
\label{aux1}
\end{equation}
while $d_1 c_1=
-\frac{\Delta^{(\epsilon_{2})}_{2}}{\Delta^{(1)}_{1}}$.
Therefore, the numerator in the right-hand side of (\ref{gamma})
is equal to
$\frac{\Delta^{(\epsilon_{i})}_{i}}{\Delta^{(0)}_{i-1}}$ and
denominator is equal to $\left (
\frac{\Delta^{(1)}_{i}}{\Delta^{(0)}_{i}} \right )^{\epsilon_i}$.
For both possible values of $\epsilon_i$, this implies that
formulae (\ref{gamma}), (\ref{cd}) and (\ref{gammadet}) agree.
\hfill$\Box$

One of the consequences of Lemma \ref{lemmaflag} is an existence
of the unique sequence of Laurent polynomials $r_0(\lambda),
\ldots, r_{n-1}(\lambda)$ of the form
\begin{equation}
r_{i}(\lambda)= \lambda^{\nu_{i+1}} + \sum_{\beta=1}^{i}
r_{i\beta} \lambda^{\nu_\beta}\ . \label{r}
\end{equation}
such that
\begin{equation}
e_1^T r_{i-1}(X) =\gamma_i e^T_i \ (i=1,\ldots,n)
\label{dualbasis}
\end{equation}
Next proposition shows that $r_i$ can be conveniently described
by formulas similar to (\ref{poly}).

\begin{prop}
\begin{equation}
r_i(\lambda)=\frac{(-1)^{i\epsilon_{i+1}}\lambda^{1 -
\sum_{\beta=1}^{i+1}\epsilon_\beta}}{\Delta^{(0)}_{i}} \P_i^{(i-
\sum_{\beta=1}^{i+1}\epsilon_\beta)}(\lambda)
\label{rat}
\end{equation}
\label{proprat}
\end{prop}

\noindent{\bf Proof.} First note, that by (\ref{interval}),
(\ref{r}) can be re-written as
\begin{equation}
r_{i}(\lambda)= \lambda^{1 -
\sum_{\beta=1}^{i+1}\epsilon_\beta}\sum_{\beta=1}^{i} a_{i\beta}
\lambda^{i}\ . \label{r1}
\end{equation}
Next, by (\ref{nu}), $\nu_{i+1}$ coincides with the lowest (resp.
highest) degree in $r_i(\lambda)$ if $\epsilon_{i+1}=1$ (resp.
$\epsilon_{i+1}=0$). Finally, due to Lemma \ref{2.1}, the claim
that $e_1^T r_i(X)$ is proportional to $e^T_{i+1}$ is equivalent
to a property
\begin{equation}{\displaystyle
e_1^T r_i(X) X^l e_1=0\ (l=0,\ldots,i-1) } \label{property}
\end{equation}
This property is satisfied by Lemma \ref{polydet}. \hfill$\Box$

\begin{cor}
Define a bilinear form $( , )$ on
$\mathbb{C}[\lambda,\lambda^{-1}]$ by
$$
(\lambda^i, \lambda^j )= s_{i+j}\ (i,j \in \mathbb{Z})\ .
$$
Then functions $p_0(\lambda),\ldots,p_{n-1}(\lambda);\
r_0(\lambda),\ldots, r_{n-1}(\lambda)$ form a bi-orthogonal system
w. r. t. $( , )$, i. e.
\begin{equation}{\displaystyle
\left (r_i(\lambda),p_{j}(\lambda)\right )= \delta_i^j c_1 d_1
\frac{\prod_{\beta=2}^{i-1}
d_\beta^{1-\epsilon_\beta}c_\beta}{(d_1\cdots
d_{i})^{\epsilon_i}}\ . } \label{bio}
\end{equation}
\end{cor}

\noindent{\bf Proof.} Follows immediately from Lemmas \ref{2.1},
\ref{lemmaflag} and Propositions \ref{3-term}, \ref{proprat}.
\hfill$\Box$

\medskip

%


\section{Symmetric Case}

It is natural to ask how should  results of the previous section
be modified, if initially one identifies $\b^*_-$ with a space
$\S$ of symmetric matrices rather than with $\H$. In this case, a
co-adjoint orbit of $\B_-$ through $X_0\in \S$ is described as
\begin{equation}
\O_{X_0}=\left \{ (\mbox{Ad}_n X_0)_{sym}\ : n\in \B_-\right \}\ ,
\label{coadsym}
\end{equation}
where $A_{sym}= A_{\ge 0} + (A_{>0})^T$.

A set $M_I$ is still defined by (\ref{M_I}, but the definition of
$X_I$ should be changed as follows:
\begin{equation}
{\displaystyle X_I=e_{1 i_1} + e_{i_1 1} + \sum_{j=1}^{k-1}(e_{i_j
i_{j+1}} + e_{i_{j+1} i_j})\ . } \label{X_Isym}
\end{equation}
An open dense subset $M'_I\subset M_I$, that we are going to
study, consists of elements of the form
\begin{equation}
{\displaystyle X=(\one - U^T_2)^{-1}(\one + U^T_1) D (\one + U_1)
(\one - U_2)^{-1}, } \label{factorsym}
\end{equation}
where, as before, $D=\mbox{diag}(d_1,\ldots,d_2)$ and $U_1, U_2$
are defined by (\ref{U12}). Matrix entries of $X$ then are found
to be
\begin{equation}{\displaystyle
x_{lm}=x_{ml}=\left \{ \begin{array}{cc}
{\displaystyle c_l\cdots c_{m-1}
 (d_l + \sum_{\alpha=i_{j-1}}^{l-1} d_\alpha (c_\alpha\cdots c_{l-1})^2 )},
&\ i_{j-1} < l \le m \le i_j\\
d_{i_{j-1}}c_{i_{j-1}}\cdots c_{m-1}, &\ i_{j-1} = l < m \le i_j\\
0, & \ \mbox{otherwise}
\end{array} \right .
} \label{xlmsym}
\end{equation}

A similar expression can be obtained for matrix entries of $X^{-1}$.
Denote by $J$ a set of indices
$\left (\{ 1,\ldots, n\}\setminus I\right ) \cup \{n \}=\{l_0 =1 < l_1 < \cdots < l_{n-k}=n \}$.
Then
\begin{equation}{\displaystyle
(X^{-1})_{lm}=(X^{-1})_{ml}=\left \{ \begin{array}{cc}{\displaystyle
(-1)^{m-l} c_l\cdots c_{m-1}
(\frac{1}{d_m} + \sum_{\alpha=m}^{l_j-1} \frac{1}{d_{\alpha+1}} (c_m\cdots c_\alpha)^2 )},
&\ l_{j-1} \le l \le m < l_j\\
(-1)^{m-l} c_{l}\cdots c_{l_j-1} \frac{1}{d_{l_j}}  , &\ l_{j-1} = l < m = l_j\\
0, & \ \mbox{otherwise}
\end{array} \right .
} \label{Xinv}
\end{equation}

Note that a conjugation of $X$ by the matrix $\mbox{diag} (\one_i,
-\one_{n-i})$ does not change values of parameters $d_j,
j=1,\ldots, n$ and $c_j, j\ne i$, but changes $c_i$ to $-c_i$.
This conjugation also does not affect values of the moments of
$X$. Thus, in order to make a solution of the inverse problem
below unique, we shall assume that all $c_i$ are positive.

As in the previous section, we are interested in expressing
parameters $c_i, d_i$ via the moment sequence $ (s_i=s_i(X)=
e_1^T X^i e_1)_{i\in \mathbb{Z}} $ of $X$.

We start by  noting that Lemma \ref{lemmaflag} remains literally
true for matrices $X$ of the form \ref{factorsym},  which implies
an existence of the unique sequence of Laurent polynomials
$r_0(\lambda), \ldots, r_{n-1}(\lambda)$ of the form (\ref{r})
satisfying (\ref{dualbasis}). Formulas for functions
$r_i(\lambda)$ are similar to (\ref{rat}).

First, we define a new collection of Toeplitz determinants
\begin{equation}
\D^{(m)}_i=\det T^{(i + 1 + m - 2
\sum_{\beta=1}^{i}\epsilon_\beta)}_i\ (m\in \mathbb Z)\ .
\label{detsym}
\end{equation}

\begin{prop}
\begin{equation}
r_{i-1}(\lambda)=\frac{(-1)^{(i-1)\epsilon_{i}}\lambda^{1 -
\sum_{\beta=1}^{i}\epsilon_\beta}}{\D^{(0)}_{i-1}} \P^{(i-
2\sum_{\beta=1}^{i-1}\epsilon_\beta- \epsilon_i)}_{i-1}(\lambda)
\ .
\label{ratsym}
\end{equation}
\label{propratsym}
\end{prop}

\noindent{\bf Proof.} We can argue exactly as in the proof of
Proposition \ref{proprat}. Note, however, that since $X$ is
symmetric, $r_{i-1}(X)e_1=\gamma_i e_i$. This means that condition
(\ref{property}) that guarantees (\ref{dualbasis}) has to be
replaced by
$$
e_1^T r_{i-1}(X) X^{\nu_\alpha}e_1=0 \ (\alpha=1,\ldots,i-1)
$$
or, equivalently,
\begin{equation}{\displaystyle
e_1^T r_{i-1}(X) X^l e_1=0
\ (l=1- \sum_{\beta=1}^{i-1}
\epsilon_\beta,\ldots,i-1-\sum_{\beta=1}^{i-1} \epsilon_\beta) }\
. \nonumber
\end{equation}
Function $r_{i-1}$ defined by (\ref{ratsym}) satisfies this
condition for exactly the same reason that function (\ref{rat})
satisfies (\ref{property}). \hfill$\Box$

\begin{cor}
\begin{equation}
\gamma^2_{i}= (-1)^{i-1} \frac{\D^{(0)}_{i}}{\D^{(0)}_{i-1}}
\label{gammadetsym}
\end{equation}
\end{cor}

\noindent{\bf Proof.} It follows from (\ref{flag1}),
(\ref{dualbasis}) and symmetricity of $X$ that
\begin{equation}
\gamma^2_i=e_1^T r_{i-1}(X) X^{\nu_i} e_1\ .
\end{equation}
Then (\ref{gammadetsym}) follows from Lemma \ref{polydet},
(\ref{nu}) and (\ref{ratsym}). \hfill$\Box$
\medskip

\noindent{\bf Remark.}
Note that, as one would expect, in the case of the classical moment problem
($I=\{2,3,\ldots,n \}$, $X$ is tri-diagonal), a condition that ensures
that the right-hand side of (\ref{gammadetsym}) is positive
is a positive definiteness of the ~Hankel matrix $(s_{i+j-2})_{i,j=1}^n$.
Similarly, in the case $I=\{n\}$ which  relevant in the study of the peakons lattice
and was comprehensively studied in
\cite{bss}, one comes to a conclusion that the ~Hankel matrix $(s_{2-i-j})_{i,j=1}^n$
is positive definite.
\medskip

Let us consider now an $m\times m$ sub-matrix  $X_m$ of $X\in
M'_I$ obtained by deleting $(n-m)$ last rows and columns. It is
clear from (\ref{xlmsym}), that $X_m$ does not depend on
parameters $c_m,\ldots, c_{n-1}$, $d_{m+1},\ldots, d_n$.

\begin{lem} Let $s_\alpha (X_m) = e_1 X^\alpha e_1^T$. then
\begin{equation}
s_\alpha (X_m)= s_\alpha(X)\quad (2-2\sum_{\beta=1}^m
\epsilon_\beta \le \alpha \le 2m+1 -  2\sum_{\beta=1}^m
\epsilon_\beta)\ . \label{submomrange}
\end{equation}
\label{submoment}
\end{lem}

\noindent{\bf Proof.} For $l>0$, an expression for $s_l(X)$ in terms
of matrix entries of $X$ reads $s_l(X)={\displaystyle \sum_{\alpha_1,\ldots,\alpha_{l-1}}
x_{1\alpha_1} x_{\alpha_1 \alpha_2}\cdots x_{\alpha_{l-1} 1}}$. By (\ref{xlmsym}), many of the
terms in this sum are zero. Moreover, if $l < 2 k + 1$, were $k$ is
the cardinality of $I$, one can find among the non-zero terms the one
were $\max(\alpha_1,\ldots,\alpha_{l-1})$ reaches its maximum. This term is
equal to
${\displaystyle \prod_{\beta=1}^{q} x_{i_{\beta-1}, i_\beta}^2},$ if $l=2q$
and
${\displaystyle x_{i_q i_q}\prod_{\beta=1}^{q} x_{i_{\beta-1}, i_\beta}^2},$ if $l=2q+1$.
This implies that if $m > i_j$ and $ 0 <l \le 2j+1$ then the expression for $s_l(X)$ involves
only entries of $X_m$ and, therefore, $s_l(X)=s_l(X_m)$. Since the largest $j$ such that
$m > i_j$ is given by $m -  \sum_{\beta=1}^m
\epsilon_\beta$, (\ref{submomrange}) is satisfied for $\alpha = 0, \ldots, 2m+1 -  2\sum_{\beta=1}^m
\epsilon_\beta$.

Similarly, one can use (\ref{Xinv}) to conclude that if
$ q$ is the largest index such that $l_q < m$ then (i) for $\alpha,\beta\le l_q$,
$(X_m^{-1})_{\alpha \beta} = (X^{-1})_{\alpha \beta} $;
(ii) for $-2q \le  l<0$, the expression for $s_l(X)$ contains only entries
$(X^{-1})_{\alpha \beta} $ with $\alpha,\beta\le l_q$, which means that in this case
$s_l(X)$ coincides with $s_l(X_m)$. To finish the proof, it remains to notice that
$q$ is equal to $\sum_{\beta=1}^m
\epsilon_\beta -1$.
\hfill$\Box$

\begin{prop}
\begin{equation}
\det (\lambda - X_m) = \frac{1}{\D_m^{(0)}} \P^{(m+1 -
2\sum_{\beta=1}^m\epsilon_\beta )}_{m+1}(\lambda)\ .
\label{charpolydet}
\end{equation}
\label{charpoly}
\end{prop}

\noindent{\bf Proof.} Let $=\lambda^m
+\sum_{i=0}^{m-1} a_{mi} \lambda^i$. Then the Hamilton-Cayley
theorem implies
\begin{equation}
s_{\alpha+m} (X_m) + \sum_{i=0}^{m-1} a_{mi} s_{\alpha+i}(X_m)
=0\quad (\alpha\in\mathbb{Z})\ . \label{HC}
\end{equation}
By Lemma \ref{submoment}, (\ref{HC}) remains valid if we replace
$s_{\alpha+i}(X_m)$ with $s_{\alpha+i}=s_{\alpha+i}(X)$ for
$i=0,\ldots, m$, as long as $ 2-2\sum_{\beta=1}^m \epsilon_\beta
\le \alpha \le m+1 -  2\sum_{\beta=1}^m \epsilon_\beta$ . This
means that, after the right multiplication of the matrix used in
the definition (\ref{polytoep}) of $\P^{(m+1 -
2\sum_{\beta=1}^m\epsilon_\beta )}_{m+1}$ by the unipotent matrix
$(\one + \sum_{\beta=1}^{m-1}
e_{\beta+1,m+1})$, one gets a matrix of the form
$$
\left [\begin{array}{cc} T^{(m+1 -
2\sum_{\beta=1}^m\epsilon_\beta )}_{m} & 0\\
1\ \lambda\ \cdots\ \lambda^{m-1} & \det (\lambda - X_m)\end{array}\right ]
$$
and (\ref{charpolydet}) follows.
 \hfill$\Box$

It drops out immediately from (\ref{charpolydet}) and
(\ref{factorsym}) that
\begin{equation}
d_1\cdots d_m = \frac{\D_m^{(1)}}{\D_m^{(0)}} \ .
\label{d1m}
\end{equation}

Now the analogue
of Theorem \ref{invcd} for matrices (\ref{factorsym}) drops out immediately
from (\ref{d1m}), (\ref{gammadetsym}) and (\ref{gamma}).

\begin{thm}
\begin{equation}
\displaystyle{ d_i =\frac{\D^{(1)}_{i}
\D^{(0)}_{i-1}}{\D^{(0)}_{i} \D^{(0)}_{i-1}}\ ,\quad c_i =
\frac{\left (-\D^{(0)}_{i+1}
\D^{(1)}_{i-1}\right )^{\frac{1}{2}}}{ \D^{(1)}_i}\
\left(\frac{\D^{(1)}_{i+1}}{\D^{(0)}_{i+1}}\right )^{\epsilon_{i+1}}
\left(\frac{\D^{(0)}_{i-1}}{\D^{(1)}_{i-1}}\right )^{1-\epsilon_{i}}. }
\label{cdsym}
\end{equation}
\label{invcdsym}
\end{thm}

\section{Conclusion}

The results of this paper can be used to make several constructions
used in the study of integrable lattices of Toda type more
explicit. For example, in \cite{fg1} we proved that for any $I,J$  there exists
a bi-rational Poisson map from $M'_I$ to $M'_{J}$, that intertwines
the Toda flows on  $M'_I$ and $M'_{J}$. A construction we gave was
by induction and expressions for matrix entries of elements of $M'_{J}$
in terms of matrix entries of elements of $M'_{I}$ were very involved.
Now we can give explicit formulas for this map using just (\ref{moments})
and Theorem \ref{invcd}. Similarly, Theorem \ref{invcdsym} can be used
to simplify ( in the case of elementary orbits ) the construction
of the Poisson map from the Kostant-Toda to symmetric Toda flows proposed
in \cite{bg}.

A solution to the inverse problem can also be useful in avoiding blow-ups
of the Toda flows by switching from one elementary orbit to another, which,
in turn, may become important in the context of $LU$ type algorithms for
computing eigenvalues of Hessenberg matrices. A natural question that arises
in this connection is to give an intrinsic description of all sequences $(s_i)$ from
which an element of $M'_I$ can be restored for at least one $I$.

Because of possible implications in the coding
theory, it would be also interesting to study solvability of the inverse problem
over a finite field (cf. \cite{f} where the connection with the coding
theory was observed in the tri-diagonal case).

Another direction of possible investigation that we plan to pursue
in the future is an extension of our results from $M'_I$ to $M_I$
(to this end, some of the genericity assumptions will have to be relaxed)
and then to more general co-adjoint orbits. In the latter case, one has to
restore and element of the orbit from a collection of rational functions.
In the case of generic co-adjoint orbits this has been done in \cite{dlnt,s}.


\begin{thebibliography}{}

\bibitem{akh}  N. I. Akhiezer, The classical moment problem and some related questions in
analysis. Hafner Publishing Co., New York 1965.



\bibitem{bss} R. Beals, D. H. Sattinger and J. Szmigielski [1999], Multipeakons and the classical
moment problem, {\it Adv. Math} {\bf 154} (2000), 229--257.

\bibitem{b} Yu. M. Berezansky,
The integration of semi-infinite Toda chain by means of inverse spectral
problem, {\it Rep. Math. Phys.} {\bf 24} (1986), 21--47.



\bibitem{bg} A. M. Bloch, M. Gekhtman, Hamiltonian and gradient structures
in the Toda flows, {\it Journ. of Geom.  $\&$ Phys.} {\bf 27}, 230-248 (1998)

\bibitem{bf} R. Brockett, L. Faybusovich,
Toda flows, inverse spectral transform and realization theory. Systems
Control Lett. {\bf 16} (1991), 79-88.

\bibitem{dlnt} P. Deift, L. Li, T. Nanda and  C. Tomei, The Toda flow on
a generic orbit is integrable, {\it Comm. Pure \& Appl. Math.} {\bf 39} (1986)
183--232.


\bibitem{f} L. Faybusovich, On the Rutishauser's approach to eigenvalue
problems, In:  Linear algebra for control theory, 87--102, {\it
IMA Vol. Math. Appl.} {\bf 62} (1994), 87--102.

\bibitem{fg1}  L. Faybusovich,  M. I. Gekhtman, Elementary Toda orbits and
integrable lattices, {\it J. Math. Phys. } {\bf 41} (2000),
2905--2921.

\bibitem{fg2}  L. Faybusovich,  M. I. Gekhtman,
Poisson brackets on rational functions and multi-Hamiltonian
structure for integrable lattices, Phys. Lett. A {\bf 272} (2000),
236--244.





\bibitem{kmz} S. Kharchev, A. Mironov, A. Zhedanov,
Faces of relativistic Toda chain, {\it Int. J. Mod. Phys.} {\bf 12} (1997),
2675--2724.


\bibitem{moser} J. Moser,  Finitely many mass points on the line under the influence of an
exponential potential. {\it  Batelles Recontres, Springer Notes in Physics} (1974), 417-497.









\bibitem{s} S. Singer,  Doctoral Dissertation.
Courant Institute of Mathematical Sciences. New York University (1990)

\bibitem{sur3} Y. B. Suris, Integrable discretizations for lattice systems:
local equations of motion and their hamiltonian properties, {\it Rev. Math. Phys.} {\bf 11} (1999), 727--822.

\bibitem{surpri} Y. B. Suris, Private communication.

\end{thebibliography}
\end{document}